# Study of Astronomical and Geodetic Series using the Allan Variance[1]

Z. M. Malkin

Pulkovo Observatory, Russian Academy of Sciences, St. Petersburg, Russia

**Abstract**—Recently, the Allan variance (AVAR), suggested more than 40 years ago to describe the instability of frequency standards, has been used extensively to study various time series in astrometry, geodesy, and geodynamics. This method makes it possible to effectively study the characteristics of the noise component of data, such as the change of location of stations, coordinates of radio sources, etc. Moreover, AVAR may be used to study the spectral and fractal structure of this noise component. To handle unequal and multivariate observations, which are characteristic of many astronomical and geodetic applications, the author suggests appropriate AVAR modifications. A brief overview of classical and modified AVAR in astrometry and geodynamics is given.

## INTRODUCTION

The Allan variance (AVAR) is specialized statistics developed in the 1960s to investigate frequency standards [1]. In recent years, it has also been actively used for the study of various time series in astrometry and geodynamics to investigate the earth's rotation parameters, the stability of the location of radio sources, and station coordinates [3-8,12,13,15,17]. However, the application of AVAR in its original form in astrometry and geodesy is limited by two factors. Firstly, it cannot handle unequal observations that are characteristic of most actual observations. Secondly, in some cases, it is advisable to handle multidimensional data, such as station coordinates, celestial objects, and earth rotation parameters. Below, the experience of the use of the classical AVAR in astrometry and geodynamics is briefly described, a modification of this method for multidimensional and unequal time series is proposed, and examples of their use in the processing of actual data are given.

## BASIC DEFINITIONS AND PROPERTIES OF THE ALLAN VARIANCE

The AVAR was proposed by D. W. Allan [1] as evaluation of stability of frequency standards. In classical form, AVAR is introduced as follows. Suppose we have the series of measurements $y_1, y_2, \ldots y_n$ made at successive points in time. Then, AVAR may be defined as

$$\sigma^2 = \frac{1}{2(n-1)} \sum_{i=1}^{n-1}(y_i - y_{i+1})^2 . \qquad (1)$$

The Allan variance is usually abbreviated as AVAR. In practice, the Allan deviation (ADEV) is often used. Two important observations can be made in respect of AVAR. First, AVAR is not related to any physical model of a frequency standard and uses only empirical data, i.e., measurements. Thus, there is no theoretical restriction on the use of these statistics for other types of measured quantities, including astronomy and geodynamics. Secondly, AVAR makes it possible to de-

---

scribe the behavior of a frequency standard at different averaging intervals, ranging from a period equal to the interval between readings. It suffices to consider yi as generalized measurements, representing the average values of actual measurements for a certain period of time (averaging period). To emphasize this property of AVAR, it is often referred to as $\sigma^2(\tau)$, where $\tau$ is the averaging period.

However, AVAR in its classical form (1) cannot always satisfactorily describe the astronomical and geodetic measurements. The reason for this is that these measurements are often not equally accurate, and this factor is not taken into account in (1). To overcome this limitation, the author proposes a modification of AVAR for unequal measurements [12]. It is introduced as follows. Suppose we have a series of measurements $y_1, y_2, ..., y_n$ with corresponding standard deviations $s_1, s_2, ... s_n$. Then, we can determine the quantity

$$\sigma_1^2 = \frac{1}{2p} \sum_{i=1}^{n-1} p_i (y_i - y_{i+1})^2, \qquad p = \sum_{i=1}^{n-1} p_i, \qquad p_i = (s_i^2 + s_{i+1}^2)^{-1}. \qquad (2)$$

Let us introduce notations WAVAR and WADEV for weighted AVAR and ADEV, respectively. We show the difference between the classical and weighted ADEV by the actual example. Figure 1 describes a series of definitions of the elevation H of the station with their mistakes. These data show that, in practice, the difference between ADEV and WADEV may be quite significant. We may say that the weighted estimation of WADEV is more stable with respect to outliers, provided that the outlier has the greater value of the measurement error, which is usually the case for astronomical and geodesic measurements.

However, the definition (2) has some limitations in cases where the measured quantities, while formally independent, physically are the components of one and the same multidimensional vector. Such quantities include coordinates of the pole where the independently determined quantities $Xp$ and $Yp$ are components of the two-dimensional vector of the pole position. Another example is represented by the Cartesian coordinates $X$, $Y$, and $Z$ of the station, which are three-dimensional coordinates of the radius vector in the geocentric system. Two-dimensional, spherical coordinates of celestial bodies may also be considered, such as the right ascension and declination. It is desirable to be able to handle such related series of measurements together to better fit the nature of the phenomena. To do this, a modification of AVAR for multidimensional and unequal measurements may be used as proposed by the author in [12]. It is defined as follows. Suppose we have a series of $k$-dimensional measurements $y_i = (y_i^1, y_i^2, \ldots, y_i^k)$, $i = 1, \ldots, n$, with corresponding standard deviations $s_i = (s_i^1, s_i^2, \ldots, s_i^k)$. Then, we can introduce the following estimate of AVAR

$$\sigma_2^2 = \frac{1}{2p} \sum_{i=1}^{n-1} p_i d_i^2, \qquad d_i = |y_i - y_{i+1}|, \qquad p = \sum_{i=1}^{n-1} p_i, \qquad (3)$$

where | | denotes the Euclidean norm of the vector $d_i$, representing the distance between the measured quantities in the k-dimensional space. Strictly speaking, the weight of $p_i$ should be calculated on the basis of the law of error propagation

$$p_i = \left( \sum_{j=1}^{k} \left\{ \left[ (y_i^j - y_{i+1}^j)/d_i \right]^2 \left[ (s_i^j)^2 + (s_{i+1}^j)^2 \right] \right\} \right)^{-1}. \qquad (4)$$

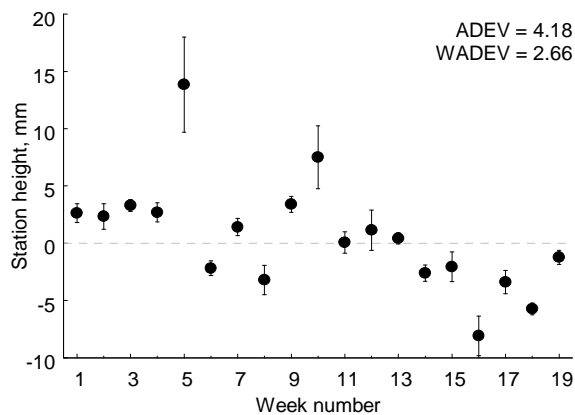

**Fig. 1.** Example of station height time series.

However, this formula has a singularity near $d_i = 0$, i.e., in the case of two equal or near-equal consecutive measurements. After a series of experiments to calculate the weights, the following simplified empirical expression was adopted

$$p_i = \left( \sum_{j=1}^{k} \left[ (s_i^j)^2 + (s_{i+1}^j)^2 \right] \right)^{-1}. \qquad (5)$$

Test results of the treatment of various series of measurements showed nearly the same practical value of (4) and (5). Let us denote weighted and multidimensional AVAR and ADEV as WMAVAR and WMADEV, respectively.

The AVAR is a characteristic of noise (random) component of the measured signal. Theoretical analysis and results of practical application make it possible to determine its main differences from other estimates of the noise component, primarily from the most widely used root mean square deviation from the mean. It is easy to see that the value of ADEV, in contrast to the root mean square deviation, is almost independent of long-period variations and trends in the studied process, as well as from abrupt changes in the measured quantity (when the number of jumps is much smaller than the number of measurements).

In [5, 9], it is recommended to use AVAR to study the spectral properties of noise in the signal under the assumption that its spectral density may be described by a power law of the following form

$$S(f) = S_0 \left( \frac{f}{f_0} \right)^k, \qquad (6)$$

where $S_0$ and $f_0$ are constants. To do this, we calculate AVAR for the sequence of averaging intervals $\tau$ from a unit interval between readings up to a third of the half of the series duration. This procedure may be performed for both independent intervals and intervals with an overlap. Further, using the method of least squares, it is necessary to calculate the coefficient $\mu$ of the linear regression

$$\log(\sigma^2(\tau)) = \mu \cdot \log(\tau) + b. \qquad (7)$$

Then, the type of the noise prevailing in a series of measurements may be defined as follows

$$\mu = \frac{\log(\sigma^2(\tau))}{\log(\tau)} \quad \begin{cases} < 0 & - \text{ white noise} \\ = 0 & - \text{ flicker noise} \\ > 0 & - \text{ random walk} \end{cases} \qquad (8)$$

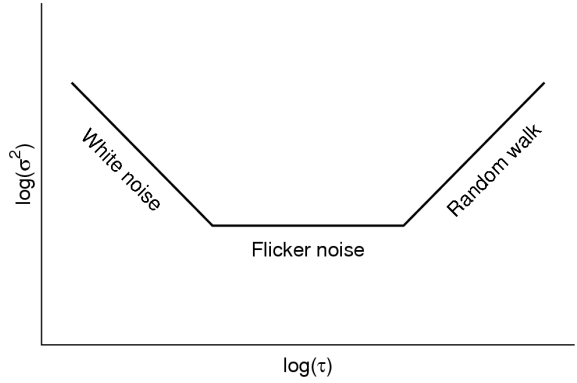

**Fig. 2.** The graphical determination of the type of the noise component of the signal by AVAR $\sigma^2$, calculated for different intervals of averaging $\tau$

In practice, it is best to build on a logarithmic scale the diagram of dependence of AVAR on $\tau$, which shows the type of the noise prevailing in the investigated signal (Fig. 2). It should be noted that study of the spectral type of the noise in the series of station coordinates is important to obtain a realistic assessment of the velocity dispersion and other related quantities [18]. Simple expressions for the station rate error, determined by measuring the presence of a certain kind of noise, are given in the works of Nicolaidis [14]. Suppose we have $n$ measurements made at intervals $\Delta T$. Also, let $a$ be the amplitude of the noise. Then, for the station velocity dispersion $s_v$ we have

$$s_v^2 = \begin{cases} \dfrac{12a^2}{\Delta T^2(n^3 - n)} & - \text{ white noise} \\ \dfrac{9a^2}{16\Delta T^2(n^2 - 1)} & - \text{ flicker noise} \\ \dfrac{a^2}{\Delta T(n - 1)} & - \text{ random walk} \end{cases} \qquad (9)$$

Williams [18] gave a derivation of the general form of the covariance matrix for a random noise signal with a spectrum described by an exponential function (6).

Finally, AVAR may be used to estimate a characteristic time series, such as the Hurst exponent [2], which allows us to estimate the mutual importance of random and trend components. Thus, the calculation of the Hurst exponent may be used for a more complete description of the properties of the investigated time series.

Despite the apparent simplicity of formulas (7)–(9), their use for the practical examination of spectral and fractal properties of real astronomical and geodetic time series is quite a challenge. There are indications that reliable results may be obtained only for series with the duration of at least several hundred of points [18]. In addition, as experience shows, in determining the noise spectral properties, seasonal and other known periodic components should be pre-excluded from the series.

**ALLAN VARIANCE APPLICATION EXAMPLES**

In recent years, AVAR has been actively used in various astronomical and geodynamic research and practical applications, such as determining the parameters of the earth's rotation and variations of coordinates of stations and radio sources. In the examples below, the first works are listed in which AVAR was applied to certain types of data and studies, along with selected later works, developing the earlier studies.

Let us give two examples of using AVAR when analyzing the parameters of the earth's rotation. For several years, AVAR was used in the International Earth Rotation and Reference Systems Service (IERS) in the process of calculating the combined series of parameters of the earth's rotation. With the help of AVAR, the series were weighted and their quality was calculated for the differences of the original series of the earth's rotation at different times of averaging. This method was introduced into the practice of IERS by Feissel [3] and was used until 2005, when the algorithm for computing the combined series of Earth rotation parameters in IERS was completely changed [8].

In [12], the author proposed the use of the AVAR to estimate the noise component of the nutation series, calculated according to the VLBI observations using different coordinate catalogs of radio sources. The meaning of this proposal is as follows. When comparing catalogs, a major problem lies in the fact that the known methods of comparison allow us to investigate only differences of errors of compared catalogs but do not allow the determination of absolute values of these errors (external precision) for each catalog. One possible way of assessing the quality of such catalogs was proposed in [12]. This method is based on the fact that errors of coordinate sources are one of the main factors influencing the error of the observed nutation angles (in other words, coordinates of the celestial pole). To estimate the noise component, the modification of AVAR for two-dimensional, unequal observations was used. In particular, the use of this method to compare the Pulkovo union catalog RSC(PUL)07C02 and ICRF showed higher accuracy of the union catalog obtained by Sokolova and Malkin [17]. Table 1 shows the values of WADEV for coordinates X and Y of the celestial pole and the two-dimensional version (2D).

Some researchers successfully used AVAR (ADEV) to analyze the series of coordinates of stations and related quantities. Malkin and Voinov first used ADEV to estimate random variations of station coordinates of the European GPS network EUREF, obtained by different processing methods, making it possible to compare the quality of these methods [13]. Later this approach was used by C.A. Roberts et al. to assess the random error of a series of base lengths, which was then used to construct a significance test of the observed changes in the lengths of bases caused by crustal deformations due to volcanic activity [15]. In works of Feissel-Vernier et al. [6, 9] one may find examples of AVAR use for studying variations in station coordinates VLBI, SLR, GPS, and DORIS, including the assessment of the spectral type of the noise component in changes of coordinates. In [7], AVAR was used for a detailed study of the geocenter motion determined by the methods of space geodesy and its comparison with geophysical models.

**Table 1.** Values of WADEV (microseconds of arc) for a series of coordinates of the celestial pole calculated with two catalog of radio source coordinates, microarcseconds.

| Каталог | $X$ | $Y$ | $2D$ |
|---|---|---|---|
| ICRF-Ext.2 | 113 | 109 | 111 |
| RSC(PUL)07C02 | 105 | 106 | 105 |

Allan variance has been also used to analyze the series of coordinates of radio sources obtained from the daily series of VLBI observations. Feissel-Vernier et al. used AVAR for the first time to study the spectral properties of the variations of radio source positions [5]. In [4], AVAR was used as a coordinate stability criterion for the choice of the defining sources for the new implementation of ICRF. In this case, Feissel-Vernier applied AVAR to average normal points [4], and Malkin [ftp://ivscc.gsfc.nasa.gov/pub/memos/ivs2009001v01.pdf] used original values of radio source coordinates obtained for daily sessions and given by analysis centers of the International VLBI Service for Geodesy and Astrometry (IVS) [14] in the framework of the ICRF2 project [10]. Each analysis center involved in this work presented one to 12 solutions containing a coordinate series of 32 to 850 selected radio sources. These solutions are available in IVS data centers [ivscc.gsfc.nasa.gov/ivsmisc/ICRF2/timeseries/]. For definiteness, we distinguish the solution as a set of series of coordinates of a variety of sources obtained by the analysis center during one cycle of treatment of VLBI observations and the series of coordinates of one source included in the decision and representing coordinates of the source at different epochs.

The analysis of a series of coordinates gives several statistic criteria for the selection of reference sources. In particular, they may be systematic changes of coordinates, both linear (the rate of apparent motion) and irregular, as well as their scatter about the mean, linear trend, or a more complex model of systematic variations (noise). It turns out that the velocity of sources, the noise component, and even jumps in coordinates depend significantly on used models of reduction, the method of processing, and software [11, ftp://ivscc.gsfc.nasa.gov/pub/memos/ivs2009001v01.pdf]. Let us cite as an example two sets of coordinates of the source 0528+134, calculated at the USNO by different methods (Fig. 3). Even the first visual comparison of these series shows how they are different in respect to noise and systematic components. More examples are given in [11] and [ftp://ivscc.gsfc.nasa.gov/pub/memos/ivs2009001v01.pdf]

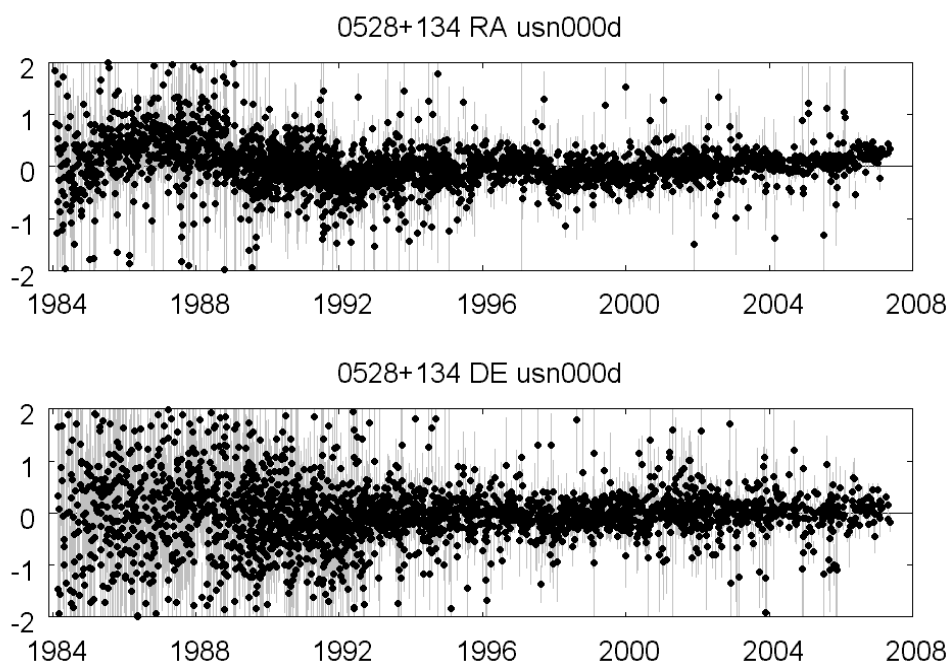

**Fig. 3.** Two series of coordinates of the radio source 0528 + 134 (deviations from the mean value) calculated at the USNO, usn000d (on the left) and usn001a (on the right).

Detailed analysis of this issue is beyond the scope of this paper. We note only that the two series were calculated with different sets of estimated parameters. In calculating the series usn000d, unlike usn00la, for each daily session of observations the day length, angle of nutation, and coordinates of stations were evaluated. Coordinates of sources were calculated in different ways. Further, we restrict ourselves to the study of the random component for a variety of sources. For this purpose, we chose 15 solutions from different analysis centers, containing the largest number of sources and sessions, as well as representing different software packages and processing methods. Then, samples from the original series were formed (sample series) for sources and sessions that are common to all initial solutions. For these sample series and for each series and source, the mean value, linear trend (rate), root mean square deviation, and ADEV were computed. In addition, for each solution, median values of these parameters were calculated for all sources as a general index of the noise inherent in the series.

Table 2 presents data relating to the noise component of coordinate series for seven sources observed most frequently. In this paper, we are interested in comparing the two estimates of the random (noise) component, i.e., the root mean square deviation (RMS) and ADEV. It is seen that, in most cases, they are quite close to each other but apparent differences are also observed. A comparison of the root mean square deviation and ADEV with graphs of changes of coordinates of sources shows that the difference between the root mean square deviation and ADEV is greater the greater the systematic change of coordinates. For example, let us compare data from series of coordinates of the source 0528+134 usn000d and usn001a. For a more detailed comparison with Fig. 3, we list values of the RMS and ADEV for each coordinate separately (Table 3).

**Table 2.** 2D estimates of the RMS (the first number) and ADEV (the second number) for the most frequently observed radio sources and median values for all sources, microarcseconds. N is the number of common to all series periods (sessions)

| Series | 0552+398 N=2168 | 0923+392 N=1919 | 1741-038 N=1868 | 0851+202 N=1678 | 0727-115 N=1675 | 0528+134 N=1653 | 1749+096 N=1516 | Median |
|---|---|---|---|---|---|---|---|---|
| bkg000c | 207 / 209 | 316 / 251 | 385 / 420 | 324 / 330 | 413 / 472 | 362 / 370 | 312 / 353 | 457 / 515 |
| dgf000f | 235 / 233 | 344 / 289 | 547 / 546 | 381 / 372 | 558 / 585 | 453 / 436 | 433 / 433 | 512 / 557 |
| dgf000g | 180 / 186 | 302 / 248 | 437 / 450 | 296 / 306 | 476 / 524 | 368 / 368 | 323 / 350 | 442 / 499 |
| gsf001a | 161 / 161 | 258 / 208 | 294 / 316 | 275 / 262 | 365 / 410 | 334 / 337 | 234 / 261 | 394 / 449 |
| gsf003a | 163 / 166 | 254 / 203 | 311 / 334 | 273 / 273 | 350 / 397 | 321 / 321 | 238 / 261 | 400 / 456 |
| iaa001b | 200 / 210 | 317 / 245 | 413 / 426 | 380 / 336 | 442 / 495 | 397 / 409 | 303 / 348 | 464 / 529 |
| iaa001c | 284 / 288 | 340 / 289 | 486 / 473 | 386 / 384 | 502 / 552 | 425 / 444 | 393 / 407 | 522 / 585 |
| mao000b | 442 / 434 | 511 / 484 | 728 / 713 | 612 / 566 | 705 / 738 | 682 / 633 | 657 / 620 | 675 / 718 |
| mao006a | 486 / 463 | 563 / 519 | 931 / 803 | 775 / 636 | 937 / 889 | 868 / 706 | 857 / 709 | 740 / 767 |
| opa000b | 355 / 338 | 435 / 393 | 742 / 665 | 519 / 473 | 669 / 671 | 606 / 526 | 650 / 563 | 577 / 588 |
| opa002a | 174 / 178 | 278 / 218 | 307 / 330 | 284 / 270 | 344 / 394 | 328 / 330 | 242 / 281 | 382 / 430 |
| sai000b | 504 / 581 | 624 / 718 | 968 / 1113 | 699 / 766 | 1146 / 1372 | 852 / 978 | 789 / 915 | 906 / 1076 |
| sha006a | 185 / 188 | 249 / 217 | 354 / 387 | 278 / 278 | 468 / 505 | 389 / 366 | 296 / 337 | 446 / 501 |
| usn000d | 167 / 172 | 279 / 219 | 331 / 354 | 289 / 281 | 379 / 432 | 338 / 346 | 260 / 295 | 402 / 468 |
| usn001a | 445 / 429 | 522 / 472 | 722 / 711 | 620 / 550 | 670 / 693 | 690 / 630 | 660 / 620 | 680 / 732 |

**Table 3.** Estimates of the RMS and ADEV for the source 0528+134, microarcseconds

| Series | $\Delta\alpha \cdot \cos\delta$ | | $\Delta\delta$ | |
|---|---|---|---|---|
| | RMS | ADEV | RMS | ADEV |
| usn000d | 205 | 185 | 253 | 277 |
| usn001a | 356 | 352 | 568 | 506 |

Table 4. Values of ADEV for Russian and Ukrainian EPN stations for the (E) east, (N) north, and (U) zenithal components of displacement of stations

| Station | Location | ADEV, mm | | |
|---|---|---|---|---|
| | | E | N | U |
| CNIV | Chernihiv, Ukraine | 2.8 | 1.6 | 2.6 |
| EVPA | Evpatoria, Ukraine | 0.7 | 0.6 | 1.9 |
| GLSV | Golosiiv, Ukraine | 1.1 | 0.8 | 2.7 |
| KHAR | Kharkiv, Ukraine | 1.6 | 0.9 | 3.2 |
| MDVJ | Mendeleevo, Russia | 1.1 | 0.9 | 4.8 |
| MIKL | Nikolaev, Ukraine | 0.7 | 0.5 | 1.9 |
| POLV | Poltava, Ukraine | 0.9 | 0.5 | 2.9 |
| PULK | Pulkovo, Russia | 0.7 | 0.6 | 2.7 |
| SULP | Lviv, Ukraine | 0.6 | 0.5 | 1.9 |
| SVTL | Svetloe, Russia | 0.7 | 0.7 | 2.4 |
| UZHL | Uzhgorod, Ukraine | 0.9 | 0.6 | 1.8 |
| ZECK | Zelenchukskaya, Russia | 0.9 | 0.8 | 2.2 |

As can be seen from Fig. 3, systematic changes of the right ascension for the two compared series are close to each other, and values of the RMS and ADEV also roughly the same. Yet we can see that the difference between the root mean square deviation and ADEV for the series usn000d is greater because of less overall noise level compared to usn001a. As a consequence, systematic changes in right ascension close by values of two series have a greater effect on the ratio of estimates of the root mean square deviation and ADEV for this series. On the other hand, in the series usn001a more significant changes are observed in the source declination, which corresponds to a significantly lower value of the ADEV compared with the root mean square deviation for this case (Table 3). As a result, we can say that practice confirms the above conclusion about the low sensitivity of AVAR to systematic changes in the studied signal and, hence, greater stability of the estimate of the quantity of the random component in comparison with the RMS. The same example shows that, to obtain estimates of the root mean square deviation not distorted by systematic changes in coordinates of the source, it is necessary to do a fairly delicate and controversial work on the approximation and elimination of the systematic component, while using ADEV this result may be achieved automatically.

The property of ADEV's small dependence on the systematic changes makes it preferable for the assessment of random variations of station coordinates. In the work of Malkin and Voinov [13] based on a comparison of the noise component in series of station coordinates of the European GPS network EUREF, the advantage was shown of the authors method of processing without fixing the coordinates of reference stations. This same method may be applied to modern observations to compare the series of station coordinates as one of possible assessments of the quality of observational data. For example, let us take Russian and Ukrainian stations participating in the European Permanent GPS Network (EPN). For the analysis, we use weekly station coordinate values, available at the EPN Central Bureau [ftp.epncb.oma.be]. To make the comparison more stringent, four Russian and eight Ukrainian GPS stations were selected operating in the period from May 2008 (the date of inclusion of the Pulkovo station PULK in the EPN) to June 2009 (GPS weeks 1480–1536). Strictly speaking, data from all stations operating at present, except for the ALCI station, which joined the EPN not so long ago, were processed. The KHAR station had an idle period of approximately 5 months, but the data from it was used in this comparison. Table 4 shows the ADEV for these stations for three components of displacement in the topocentric coordinate sys-tem. These data allow us to estimate the random component of variation of coordinates that may be an indicator of the quality of the receiver, antenna, the stability of its installation, and, possibly, other factors. Once again, we emphasize that ADEV serves as an estimate of only the random component of the station displacement, and its value has usually little correlation with the value of systematic displacement components, the main of which are usually trends and seasonal variations and, sometimes, jumps in coordinates.

## CONCLUSIONS

The Allan variance is effective and promising statistics for the study of time series of observational data. When used in addition to other traditional methods of research, it makes it possible to obtain independent data on the noise component of the studied signals, practically independent of the presence of long-period components and discontinuities, in contrast to the commonly used estimate of the mean square deviation. AVAR also allows us to study the spectral properties of the noise component more efficiently from a computational point of view than the direct calculation of the signal spectrum.

To study unequal and multidimensional data sets typical of many practical applications in astronomy, geodesy, and geodynamics, modifications of the classical AVAR proposed by the author may be used [12]. On the basis of actual data, it is shown that the weighted estimate of AVAR is more resistant to the outliers than the classical one.